\documentclass[12pt]{article}

\usepackage{amssymb}
\usepackage{amsmath}
\usepackage{amscd}
\usepackage{latexsym}
\usepackage{graphicx}

\usepackage{cite}

\newcommand{\be}{\begin{equation}}
\newcommand{\ee}{\end{equation}}

\newcommand{\dlt}{\delta}
\newcommand{\prt}{\partial}

\newcommand{\bt}{\beta}
\newcommand{\vp}{\varphi}
\newcommand{\ep}{\varepsilon}
\newcommand{\al}{\alpha}
\newcommand{\ra}{\rightarrow}

\newcommand{\Gm}{\Gamma}

\newcommand{\lbd}{\lambda}
\newcommand{\Lbd}{\Lambda}

\newcommand{\cA}{{\cal A}}
\newcommand{\cL}{{\cal L}}

\begin{document}

\begin{center}
 
{\Large{\bf Strong-coupling limits induced by weak-coupling expansions} \\ [5mm]

V.I. Yukalov$^{1,2}$ and E.P. Yukalova$^{3}$}  \\ [3mm]

{\it
$^1$Bogolubov Laboratory of Theoretical Physics, \\
Joint Institute for Nuclear Research, Dubna 141980, Russia \\ [2mm]

$^2$Instituto de Fisica de S\~ao Carlos, Universidade de S\~ao Paulo, \\
CP 369, S\~ao Carlos 13560-970, S\~ao Paulo, Brazil \\ [2mm]

$^3$Laboratory of Information Technologies, \\
Joint Institute for Nuclear Research, Dubna 141980, Russia } \\ [3mm]

{\bf corresponding author}: {\it yukalov@theor.jinr.ru} \\

\end{center}

\vskip 1cm

\begin{abstract}

A method is described for the extrapolation of perturbative expansions in powers of 
asymptotically small coupling parameters or other variables onto the region of finite 
variables and even to the variables tending to infinity. The method involves the 
combination of ideas from renormalization group theory, approximation theory, dynamical 
theory, and optimal control theory. The extrapolation is realized by means of self-similar 
factor approximants, whose control parameters can be uniquely defined. The method allows 
to find the large-variable behavior of sought functions knowing only their small-variable 
expansions. Convergence and accuracy of the method are illustrated by explicit examples, 
including the so-called zero-dimensional field theory and anharmonic oscillator. 
Strong-coupling behavior of Gell-Mann-Low functions in multicomponent field theory, 
quantum electrodynamics, and quantum chromodynamics is found, being based on their 
weak-coupling perturbative expansions.         

\end{abstract}

\vskip 2mm


\newpage

\section{Introduction}

The majority of interesting physical problems cannot be solved exactly, but succumb 
only to perturbation theory resulting in asymptotic, usually divergent, expansions. 
Such expansions have, as a rule, the form of truncated series in powers of a parameter,
which often is a coupling parameter, or some other variable assumed to be asymptotically 
small. Usually such series are divergent for any finite value of the considered variable. 
However, the realistic values of these variables are usually finite, and in many cases 
the most interesting region is that of asymptotically large variables. This is why the 
extrapolation of divergent expansions from the region of asymptotically small variables 
to that of finite variables has been the topic of intensive research. 

Several tools have been developed for such an extrapolation to the region of finite 
variables, e.g. Pad\'{e} approximants \cite{Baker_1}, Borel summation \cite{Hardy_2}, 
change of variables, such as conformal mapping \cite{Nehari_3}, and series transformations, 
for instance Schmidt-Shanks transformation \cite{Weniger_4}. These tools can allow for 
accurate extrapolation to finite variables, but they cannot be used for the extrapolation 
to asymptotically large variables.

Consider, for example, the Pad\'{e} approximants constructed from a perturbative expansion 
of order $k$ in powers of a variable $x$, 
$$
P_{M/N}(x) = \frac{a_0 + a_1 x + \ldots + a_M x^M}{1 + b_1 x + \ldots + b_N x^N}   ,
$$  
where $M+N=k$. It can provide a reasonable approximation for finite $x$, but at 
asymptotically large $x$, it behaves as
$$
P_{M/N}(x) \simeq B_{M/N} \; x^{\nu_{M/N}} \qquad (x \ra \infty) \;   ,
$$
with the large-variable amplitude and exponent
$$
B_{M/N} = \frac{a_M}{b_N} \; , \qquad  \nu_{M/N} = k - 2N  \qquad 
( 0 \leq N \leq k) \;  .
$$
Since, for each order $k$, there exists the whole table of approximants for different $M$ 
and $N$, the large-variable limit is not defined giving, depending on $M$, $N$, and fixed 
$k$, the set of large-variable exponents
$$
\nu_{M/N} = - k, ~ - k + 2 , ~ - k +4 , \ldots , ~ k-4, ~ k - 2 , ~ k \;   ,
$$
which varies with $k$. 
   
Series transformations are usually composed of power-law expansions, thus being reduced to 
the ratio of polynomials, similarly to Pad\'{e} approximants. Hence, in the same way as for
the latter, the limit $x \ra \infty$ is not defined, although for finite $x$ a reasonable 
accuracy can be achieved. 
   
When using some change of variables, one makes a mapping $x=x(z)$, like conformal mapping,
and re-expands the given truncated series in powers of the new variable $z$. Then one 
again confronts the necessity of resorting to an effective summation of the new series, 
with returning back to the old variable $x$. Again, such a procedure can lead to a 
reasonable approximation for finite $x$, but it is not uniquely defined for $x$ tending 
to infinity, being dependent on the particular mapping and the summation method for the 
series in terms of new variables.   

Resorting to the Borel summation, one needs to find an effective sum for the Borel 
transform
$$
B_k(x) = \sum_{n=0}^k \frac{a_n}{n!} \; x^n \;   ,   
$$
say by using Pad\'{e} approximations, $B_{M/N}(x)$, and to substitute it in the inverse 
Borel transformation. For example, in the widely employed Pad\'{e}-Borel method, one has
$$
f_{PB}(x) = \int_0^\infty B_{M/N}(xt) \; e^{-t} \; dt \;   .
$$
For large $x$, Pad\'{e} approximants behave as
$$
 B_{M/N}(x) \simeq C_{M/N} x^{M-N} \qquad (x \ra \infty) \;  ,
$$
so that for the sought function, one has the large-variable form
$$
f_{PB}(x) \simeq  C_{M/N}\; \Gm(M - N + 1)\; x^{M-N} \qquad (x \ra \infty) \;  ,
$$
which is not uniquely defined, analogously to the case of Pad\'{e} approximants.

In this way, it is clear that the existing methods of defining effective sums of 
divergent perturbative expansions, although can yield good approximations for finite 
values of variables, but are not able to correctly characterize the limit of 
asymptotically large variables.    

In the present paper, we describe a method allowing for the extrapolation of perturbative 
expansions, derived for asymptotically small variables, to the region of finite and even
infinite variables. The sequence of steps is as follows. First, we present the justification 
of the method. Then the approach is applied to the problems with known solutions, so that
the convergence and accuracy of the approach could be explicitly illustrated. After that,
predictions are made for the problems whose large-variable behavior is not known. This
concerns the strong-coupling limit of the Gell-Mann-Low functions in multicomponent 
$\varphi^4$ field theory, quantum electrodynamics and quantum chromodynamics.

\section{Controlled approximation cascade}

Let us consider a complicated problem whose solution can be found only in the region
of an asymptotically small parameter or variable, where perturbation theory yields the 
asymptotic truncated series
\be
\label{1} 
f_k(x) = \sum_{n=0}^k a_n x^n \qquad ( x \ra 0 ) \;   ,
\ee
where, without the loss of generality, we can set $a_0 = 1$. The more general case, where
$$
f_k(x) = f_0(x) \sum_{n=0}^k a_n x^n   ,
$$
with a known function $f_0(x)$, can be easily reduced to the above form by considering 
$f_k(x)/f_0(x)$. For concreteness, we keep in mind a real function of a real variable 
$x \in [0,\infty)$.

Suppose our problem of the most interest is to find the large-variable behavior of a 
function knowing solely its small-variable expansion (\ref{1}). Can we find the behavior
of a function $f(x)$ at $x \ra \infty$ knowing only its behavior at $x \ra 0$?  

The pivotal idea of the approach, we are presenting, is to reformulate the sequence of 
perturbative truncated series into a dynamical system, with the approximation order $k$
playing the role of discrete time \cite{Yukalov_5,Yukalov_6,Yukalov_7,Yukalov_8}. 
If we are able to accomplish this, then the equation of motion for the dynamical system 
can be analyzed for the purpose of finding a fixed point representing an effective limit 
of the series. To our understanding, although expansion (\ref{1}) is obtained for 
$x \ra 0$, but its coefficients $a_n$ contain information on the whole function 
representing its effective limit, similarly to how the coefficients of a Taylor expansion 
contain information of the whole function which the expansion is obtained from. The goal 
is to extract the information hidden in the coefficients $a_n$.    
 
Keeping in mind that the sequence of expansions (\ref{1}) is usually divergent, it is clear 
that by itself it cannot form a stable dynamical system. In order to stabilize a dynamical 
system, one needs to introduce control functions that would transform the system so that
to govern its stability and provide the existence of a fixed point, similarly to how this
is done in optimal control theory \cite{Bellman_9,Lee_10}. 

The transformation of expansions (\ref{1}), incorporating control functions, can be 
schematically represented as an operation 
\be
\label{2}
 F_k(x,u_k) = \hat T[\; u_k\; ] \; f_k(x) \;  .
\ee
Control functions $u_k = u_k(x)$ are to be defined so that to provide convergence to the 
renormalized sequence of the approximants $F_k(x,u_k)$, which implies the validity of the 
Cauchy criterion, when for any $\varepsilon > 0$ there exists $k_\varepsilon$ such that
\be
\label{3}
 | \; F_{k+p}(x,u_{k+p}) - F_k(x,u_k) \; | < \ep  
\ee
for $k > k_\varepsilon$ and $p > 0$. 

The renormalized sequence $\{F_k(x,u_k)\}$ does not compose yet a dynamical system. 
For the correct mathematical definition of a dynamical system, one has to construct 
a phase space, or state space, define an endomorphism acting on that space, and formulate 
the evolution law \cite{Katok_11}. For this purpose, we impose a rheonomic constraint
\be
\label{4}
f = F_0( x,u_k(x) ) \; , \qquad x = x_k(f)
\ee
and introduce the function
\be
\label{5}
 y_k(f) = F_k(x_k(f) , u_k( x_k(f) ) ) \;  .
\ee

The sequences $\{y_k(f)\}$ and $\{F_k(x,u_k)\}$ are bijective by construction. The closed
linear envelope 
\be
\label{6}
\cA = \overline\cL\{ y_k(f) : ~ k = 0,1,2,\ldots \}
\ee
constitutes the {\it approximation space} \cite{Pietsch_12}. The function $y_k(f)$ is the 
sought endomorphism acting on the approximation space. The dynamical system in discrete time
\be
\label{7}
\{ y_k(f) : ~ \mathbb{Z}_+ \times \cA ~\longmapsto ~ \cA \} 
\ee
is called cascade. In the present case this is an {\it approximation cascade}, whose 
trajectory
\be
\label{8} 
y_k(f) ~ \longmapsto ~ y_{k+1}(f) ~ \longmapsto ~ \ldots ~ \longmapsto ~ y^*(f)
\ee
is bijective to the sequence of the approximants $\{F_k(x,u_k)\}$.   

Finally, we have to formulate the evolution equation for the approximation cascade.
The Cauchy criterion now takes the form
\be
\label{9}
 | \; y_{k+p}(f) - y_k(f) \; | < \ep     .
\ee
By assumption, the control functions guarantee the existence of a limit
\be
\label{10}
  y_{k+p}(f) \simeq y^*(f) \;  ,
\ee
when $k$ increases. This limit corresponds to a fixed point that, by definition, satisfies
the equation
\be
\label{11}
 y_k(y^*(f)) = y^*(f)) \;  .
\ee
From these equations, it follows that in the vicinity of a fixed point the endomorphism
obeys the evolution equation
\be
\label{12}
y_{k+p}(f) = y_k(y_p(f)) \; 
\ee
that can be called {\it self-similar relation} \cite{Yukalov_5,Yukalov_6,Yukalov_7,Yukalov_8}.     
This is a group property typical of the theory of dynamical systems as well as of 
renormalization group theory \cite{Bogolubov_13,Yukalov_14}. 

In this way, the approximation cascade (\ref{7}) is defined, with the approximation  
space (\ref{6}), endomorphism (\ref{5}), the self-similar evolution equation (\ref{12}),
and the approximation order $k$ playing the role of discrete time. Since in practical 
calculations, we cannot reach the limit of $k \ra \infty$, but have to deal with a finite
approximation order $k$, the approximation for the sought fixed point will depend on the 
considered order $k$, being denoted as $y_k^*(f)$.  

A fixed point of the cascade, $y^*_k(f)$, by construction, is bijective to the effective 
limit of the perturbative sequence called the self-similar approximation:
\be
\label{13}
 f_k^*(x) = \hat T^{-1} [\; u_k \; ] \; y_k^*(F_0(x,u_k(x))) \;  .   
\ee
The details of techniques for determining fixed points of approximation cascades can 
be found in Refs. \cite{Yukalov_14,Yukalov_15,Yukalov_16}.

\section{Self-similar factor approximants}

The practical realization of the above program can be accomplished as follows. Control 
functions can be incorporated into the considered sequence in several ways. Since 
the main idea is to extract information from the given sequence of expansions in a 
self-similar manner, we use the {\it fractal transform}
\be
\label{14}
 F_k(x,\{ n_j\} ) = \prod_{j=1}^k x^{-n_j} \; f_k(x) \; ,
\ee
with the inverse transformation
\be
\label{15}
f_k(x) = \prod_{j=1}^k x^{n_j} \; F_k(x,\{ n_j\} ) \;   .
\ee
Expansion (\ref{1}) can be rewritten as
\be
\label{16}
f_k(x) = \prod_{j=1}^k ( 1 + b_j x) \;   ,
\ee
with $b_j$ expressed through $a_n$. Then the fractal transform (\ref{14}) can be 
represented in the form
\be
\label{17}
F_k(x,\{ n_j\}) = \prod_{j=1}^k x^{-n_j} \; ( 1 + b_j x) \;  .
\ee

Following the scheme described in the previous section, we obtain 
\cite{Gluzman_17,Yukalov_18}
the fixed point of order $k$ giving the self-similar factor approximant 
\be
\label{18}
 f_k^*(x) = \prod_{j=1}^{N_k} ( 1 + A_j x )^{n_j} \;   ,
\ee
where, depending on whether the approximation order is even or odd,
\begin{eqnarray}
\label{19}
N_k = \left\{ \begin{array}{ll}
k/2 , ~ & ~ k = 2,4, \ldots \\
(k+1)/2  ~ & ~ k = 3,5, \ldots
\end{array}
\right. \; .
\end{eqnarray}
  
In order to define explicit forms of the control quantities $A_j$ and $n_j$, it 
is necessary to impose some optimization conditions. From the Cauchy criterion of 
convergence, it is straightforward to derive the optimization conditions in the 
form of minimal-difference and minimal-derivative conditions 
\cite{Yukalov_14,Yukalov_15,Yukalov_16} by comparing the approximants (\ref{18}) 
of different orders and by minimizing some cost functionals, as is customary in 
determining control functions for controlled dynamical systems 
\cite{Bellman_17,Lee_18}. Deriving such general forms of the control functions 
$A_j(x)$ and $n_j(x)$ in the present case is rather expensive requiring to deal 
with several optimization conditions and making the overall use of the approach 
complicated. Instead of trying to derive $A_j(x)$ and $n_j(x)$ for arbitrary $x$, 
it is possible to resort to the learning techniques employed in machine learning
\cite{Murphy_19,Alpaydin_20}. Then one trains the considered system on the set of 
known data by comparing the approximate expressions characterizing the system with 
the given empirical data. In the present case, the known data are given by the 
explicit expansion $f_k(x)$ at small $x$. Then we need to compare the approximants 
(\ref{18}) with $f_k(x)$ by considering the minimal-difference condition
\be
\label{20}
f_k^*(x) - f_k(x) \simeq 0 \qquad ( x \ra 0 ) \; .
\ee     
The advantage of this training procedure is that the control quantities $A_j$ and 
$n_j$ become control parameters, but not functions. If the asymptotic form of the 
approximant $f_k^*(x)$ is close to expansion (\ref{1}), then the training conditions 
are
\be
\label{21}
 \lim_{x\ra 0} \; \frac{1}{n!} \; \frac{d^n}{dx^n} \; f_k^*(x) = a_n \qquad
( n = 0,1,2,\ldots) \;  .  
\ee
Explicitly, this leads to the equations
\be
\label{22}
\sum_{j=1}^{N_k} n_j A_j^m = \Lbd_m \qquad ( m = 1,2,\ldots ,k) \;    ,
\ee
where
$$
\Lbd_m = \frac{(-1)^{m-1}}{(m-1)!} \; \lim_{x\ra 0} \; 
\frac{d^m}{dx^m} \; \ln \left( \sum_{n=0}^m a_n x^n \right) \;   .
$$
Some general mathematical properties of the self-similar factor approximants (\ref{18})
are described in Appendix. 

However, equations (\ref{22}) uniquely define all control parameters $A_j$ and $n_j$
only for even orders of $k$. This is because for even $k=2,4,\ldots$, when $N_k = k/2$, 
there are $k/2$ parameters $A_j$ and $k/2$ parameters $n_j$, hence $k$ unknowns, all 
of which are defined by $k$ equations (\ref{22}). Unfortunately, for odd orders 
$k=3,5,\ldots$, when $N_k = (k+1)/2$, there are again $k$ equations (\ref{22}), but
$(k+1)/2$ parameters $A_j$ and $(k+1)/2$ parameters $n_j$, so that in total $k+1$ 
unknowns, which is larger by one than the number of equations. To uniquely define 
all control parameters for odd orders $k$, it is necessary to impose an additional 
constraint, for example, by fixing one of the parameters $A_j$, which, actually, is
a not well justified ad hoc procedure \cite{Yukalova_21}. Otherwise, one is forced to 
consider only even approximants \cite{Yukalov_22} neglecting at all odd orders.  

In the present paper, we suggest a way of overcoming the problem of odd approximation 
orders. To check the accuracy of the suggested method, we consider the most difficult
challenge of the extrapolation problem, when one needs to evaluate the large-variable 
limit of the sought function. In the following section, we formulate the method of 
dealing with odd orders.

\section{Diff-log transformation}

As is explained above, in odd orders the number of the sought parameters is larger by one 
then the number of training equations, which requires to impose an additional constraint
for uniquely defining all control parameters. This can be done as follows.

Let us define the diff-log transformation
\be
\label{23}
 D[ \; f(x)\; ] \equiv \frac{d}{dx} \;\ln f(x) \; .
\ee 
For a function with a boundary condition $f(0) = 1$, the inverse transformation is
\be
\label{24}
 f(x) = \exp\left\{ \int_0^x   D[ \; f(x)\; ]  \; dx \right\} \; .
\ee
If the large-variable limit of the function is
\be
\label{25}
f(x) \simeq B x^\nu \qquad ( x \ra \infty) \;   ,
\ee
then the large-variable exponent can be found from the limit
\be
\label{26}
\nu = \lim_{x\ra\infty} x \; D[ \; f(x)\; ] \;   .
\ee

Accomplishing for the considered truncated series (\ref{1}) the diff-log transformation
\be
\label{27}
  D[ \; f_k(x)\; ] =\frac{d}{dx} \;\ln f_k(x) \;  ,
\ee
we expand the latter in powers of $x$ getting
\be
\label{28}
 D_k(x) = \sum_{n=0}^k c_n x^n \qquad ( x \ra 0 ) \;  ,
\ee
with $b_0=a_1$. For this expansion, a self-similar factor approximant of odd order 
reads as
\be  
\label{29}
D_k^*(x) = a_1 \prod_{j=1}^{(k+1)/2} ( 1 + M_j x)^{m_j} \;   ,
\ee
with the control parameters $M_j$ and $m_j$ satisfying the system of equations
\be
\label{30}
 \sum_{j=1}^{(k+1)/2} m_j \; M_j^p =
\frac{(-1)^{p-1}}{(p-1)!} \; \left[ \;
\frac{d^p}{dx^p} \; \ln\left( \sum_{n=0}^p c_n x^n \right) \;\right]_{x=0} \;  ,
\ee
where $p=1,2,\ldots,k$. To uniquely define all control parameters for odd orders $k$, 
we have to impose an additional constraint.  

By condition (\ref{26}), the self-similar approximation for the exponent is
\be
\label{31}
\nu_k = \lim_{x\ra\infty}x \; D_k^*(x) \; ,
\ee
which requires that the approximant (\ref{29}), in the large-variable limit, behaves 
as
\be
\label{32}
 D_k^*(x) \simeq D_k x^{-1} \qquad ( x \ra \infty) \;  ,
\ee
with the additional constraint
\be
\label{33}
 \sum_{j=1}^{(k+1)/2} m_j = - 1 \;  .
\ee   
This allows to uniquely define the amplitude $D_k$, hence the large-variable exponent
\be
\label{34}
\nu_k = D_k = a_1 \prod_{j=1}^{(k+1)/2}M_j^{m_j} \; .
\ee

The found exponent $\nu_k$ can be used as a fixed constraint for uniquely defining the 
parameters $A_j$ and $n_j$ for odd orders of $k$ in equation (\ref{22}).

\section{Self-similar Borel summation}

When the coefficients of the given expansion (\ref{1}) are growing fast, it is known 
that Borel summation and its variants can help for defining an effective limit of a 
series, provided that the Borel transform can be efficiently summed. As is mentioned 
in the Introduction, the Pad\'{e}-Borel summation is not appropriate for finding the 
limiting behavior at $x \ra \infty$. However, we can employ the self-similar factor 
approximants for summing the Borel transform, combining this with the Borel transformation. 
This way, for short, will be called self-similar Borel summation. 
        
For example, the given expansion (\ref{1}) can be subject to the Borel-Leroy transformation
\be  
\label{35}
B_k(x) = \sum_{n=0}^k \frac{a_n x^n}{\Gm(n+1+u)} \;   ,
\ee 
where $u$ is an additional control parameter. For $u=0$, we have the standard Borel 
transform that we shall use. For the truncated series (\ref{35}), we can construct 
self-similar factor approximants
\be
\label{36}
B_k^*(x) = \frac{1}{\Gm(1+u)} \prod_{j=1}^{N_k} ( 1 + A_j x )^{n_j} \; ,
\ee
where the control parameters $A_j$ and $n_j$ are defined in the same way as explained 
above, but for the expansion (\ref{35}), so that $A_j$ and $n_j$ here are different 
from those in the approximant (\ref{18}).

Applying the inverse Borel transformation gives the approximant
\be
\label{37}
f_k^*(x) = \int_0^\infty e^{-t} \; t^u \; B_k^*(xt) \; dt \;   .
\ee
With the large-variable limit 
\be
\label{38}
B_k^*(x) \simeq C_k x^{\nu_k} \qquad ( x \ra \infty)
\ee
for the self-similar approximant (\ref{36}), where the amplitude is
\be
\label{39}
 C_k = \frac{1}{\Gm(1+u)} \prod_{j=1}^{N_k}   A_j^{n_j}
\ee
and the large-variable exponent is
\be
\label{40}
\nu_k = \sum_{j=1}^{N_k} n_j \;   ,
\ee
we obtain the large-variable limit for the function of interest,
\be
\label{41}
 f_k^*(x) \simeq B_k x^{\nu_k} \qquad ( x \ra \infty) \;  ,
\ee
with the amplitude
\be
\label{42}
B_k = C_k \Gm( 1 + u + \nu_k) = \frac{\Gm(1+u+\nu_k)}{\Gm(1+u)}
\prod_{j=1}^{N_k} A_j^{n_j} \;   .
\ee

For even orders of $k$, all control parameters are uniquely defined. And for odd orders 
of $k$, we follow the same trick as above, by accomplishing the diff-log transformation
\be
\label{43}
 D[\;B_k(x) \; ] \equiv   \frac{d}{dx} \; \ln B_k(x)
\ee
for the Borel transform (\ref{35}), constructing its factor approximant $D_k^*(x)$, and 
finding the large-variable exponent $\nu_k$, which serves as an additional constraint 
for uniquely determining all control parameters $A_j$ and $n_j$.

\section{Zero-dimensional $\varphi^4$ theory}

In this and the following section, we apply the developed method to the test problems, 
whose large-variable behavior is known. This will explicitly illustrate the efficiency
of the method and its convergence properties. Such an illustration is necessary before
applying the method to the problems whose large-variable behavior is not available. 

Let us start with the consideration of the generating functional 
\be
\label{44}
 Z(g) = \frac{1}{\sqrt{\pi}} 
\int_{-\infty}^\infty \exp( -\vp^2 - g \vp^4 ) \; d\vp 
\ee
of the so-called zero-dimensional $\varphi^4$ theory, which serves as a touch-stone 
for checking almost any novel approach. Expanding the integrand in powers of the 
coupling $g \ra 0$ yields the expansion
\be
\label{45}
Z_k(g) = \sum_{n=0}^k a_n g^n \qquad ( g \ra 0 ) \;   ,
\ee
with the coefficients
\be
\label{46}
a_n = \frac{(-1)^n}{\sqrt{\pi}\; n!} \;\Gm\left( 2n + \frac{1}{2}\right) \;  .
\ee
This expansion diverges for any finite value of $g$. 

Our aim is to predict, by employing the developed approach, the large-variable limiting 
behavior 
\be
\label{47}
 Z_k^*(g) \simeq B_k g^{\nu_k} \qquad ( g \ra \infty )
\ee
and to compare our predictions with the known exact numerical limit
\be
\label{48}
 Z(g) \simeq 1.022765 \; g^{-1/4} \qquad  ( g \ra \infty ) \; .
\ee
 
As is described in the previous sections, we employ two variants of the method, first, 
by directly constructing self-similar factor approximants and, second, by resorting to
self-similar Borel summation. The results are shown in Table 1 and Table 2, where the
large-variable amplitudes $B_k$ and exponents $\nu_k$ are given, with the corresponding 
percentage errors. As is seen, knowing only the small-variable expansion for $g\ra 0$,
we are able to find the large-variable behavior for $g \ra \infty$. The use of the
self-similar Borel summation improves the accuracy of the predicted results. The accuracy
is to be accepted as rather good, if we remember that the most difficult case is 
considered, when the strong-coupling behavior for $g \ra \infty$ is found being based 
solely on the knowledge of several terms of weak-coupling expansion.   

\begin{table}[hp]
\centering
\renewcommand{\arraystretch}{1.2}
\begin{tabular}{|l|c|c|c|c|} \hline
$k$  &  $B_k$  &  $\ep(B_k)\%$  &  $\nu_k$   &  $\ep(\nu_k)\%$    \\ \hline
2    &  0.823  &  -19.5         &    -0.094  &   -62.5  \\ 
3    &  0.805  &  -21.3         &    -0.137  &   -45.3   \\ 
4    &  0.806  &  -21.2         &    -0.129  &   -48.4  \\ 
5    &  0.808  &  -21.0         &    -0.159  &   -36.6   \\
6    &  0.806  &  -21.2         &    -0.148  &   -40.6   \\ 
7    &  0.815  &  -20.3         &    -0.172  &   -31.2   \\
8    &  0.810  &  -20.8         &    -0.161  &   -35.6  \\
9    &  0.823  &  -19.6         &    -0.181  &   -27.5    \\ 
10   &  0.814  &  -20.4         &    -0.170  &   -32.0  \\ 
11   &  0.830  &  -18.9         &    -0.188  &   -24.7  \\  
12   &  0.819  &  -19.9         &    -0.177  &   -29.3  \\ 
13   &  0.836  &  -18.3         &    -0.193  &   -22.6  \\ 
14   &  0.824  &  -19.4         &    -0.182  &   -27.1 \\ \hline
\end{tabular}
\caption{\small
Zero-dimensional $\vp^4$ theory. Strong-coupling amplitudes and exponents, with the 
related errors in percents, predicted by self-similar factor approximants.}
\end{table}

\begin{table}[hp]
\centering
\renewcommand{\arraystretch}{1.2}
\begin{tabular}{|l|c|c|c|c|} \hline
$k$  &  $B_k$  &  $\ep(B_k)\%$  &  $\nu_k$   &  $\ep(\nu_k)\%$    \\ \hline
2    &  0.898  &  -12.2         &    -0.207  &   -17.2  \\ 
3    &  0.936  &  -8.53         &    -0.231  &   -7.47   \\ 
4    &  0.924  &  -9.62         &    -0.226  &   -9.73  \\ 
5    &  0.955  &  -6.63         &    -0.238  &   -4.86   \\
6    &  0.941  &  -7.96         &    -0.233  &   -6.80   \\ 
7    &  0.967  &  -5.43         &    -0.241  &   -3.56   \\
8    &  0.953  &  -6.83         &    -0.237  &   -5.23  \\
10   &  0.961  &  -6.00         &    -0.239  &   -4.26    \\ 
11   &  0.978  &  -4.39         &    -0.243  &   -2.62  \\ 
12   &  0.968  &  -5.38         &    -0.241  &   -3.59  \\  
13   &  0.983  &  -3.87         &    -0.246  &   -2.19  \\ 
14   &  0.973  &  -4.88         &    -0.242  &   -3.10 \\ \hline
\end{tabular}
\caption{\small
Zero-dimensional $\vp^4$ theory. Strong-coupling amplitudes and exponents, with the 
related errors in percents, predicted by self-similar Borel summation.}
\end{table}

\section{Anharmonic oscillator}

The other touch-stone example is the anharmonic oscillator with the Hamiltonian
\be
\label{49}
 H = - \;\frac{1}{2}\;\frac{d^2}{dx^2} + \frac{1}{2} \; x^2 + g x^4 \;  ,
\ee
in which $x \in (-\infty, \infty)$ and $g \geq 0$. The ground-state energy of the 
oscillator can be expressed \cite{Bender_23,Hioe_24} in the form of the weak-coupling 
expansion
\be
\label{50}
 E_k(g) = \sum_{n=0}^k a_n g^n \qquad ( g \ra 0) \; ,
\ee
with the coefficients shown in Table 3. This series is also divergent for any finite 
$g$.

\begin{table}[hp]
\centering
\renewcommand{\arraystretch}{1.2}
\begin{tabular}{|l|l|} \hline
$n$  &  $a_n$     \\ \hline
0    &  1/2=0.5   \\
1    &  3/4=0.75  \\
2    & -0.2625$\times 10$   \\ 
3    &  0.208125$\times 10^2$  \\ 
4    &  -0.2412890625$\times 10^3$  \\ 
5    &   0.358098046875$\times 10^4$  \\
6    &  -0.639828134766$\times 10^5$  \\ 
7    &   0.132973372705$\times 10^7$   \\
8    &  -0.314482146928$\times 10^8$  \\
9    &   0.833541603263$\times 10^9$      \\
10   &  -0.244789407028$\times 10^{11}$   \\ 
11   &   0.789333316003$\times 10^{12}$  \\ 
12   &  -0.277387769635$\times 10^{14}$  \\  
13   &   0.105564665831$\times 10^{16}$\\ 
14   &  -0.432681068354$\times 10^{17}$\\ \hline
\end{tabular}
\caption{\small
Anharmonic oscillator. Coefficients of weak-coupling expansion for the ground-state
energy.}
\end{table}

Our aim is to predict, being based solely on this expansion, the strong-coupling 
behavior 
\be
\label{51}
 E_k^*(g) \simeq B_k g^{\nu_k} \qquad ( g \ra \infty)    
\ee
and to compare it with the known exact limit
\be
\label{52}
 E_k(g) \simeq 0.667986\; g^{1/3} \qquad ( g \ra \infty) \;   .
\ee
The results are given in Table 4 for the direct self-similar factor approximants and 
in Table 5 for the self-similar Borel summation. As we see, the Borel summation improves 
the accuracy by an order. Again we have to accept that even directly applying self-similar 
factor approximants, we get not bad accuracy, if to keep in mind that solely several 
terms of the weak-coupling expansion are used. 

\begin{table}[hp]
\centering
\renewcommand{\arraystretch}{1.2}
\begin{tabular}{|l|c|c|c|c|} \hline
$k$  &  $B_k$  &  $\ep(B_k)\%$  &  $\nu_k$   &  $\ep(\nu_k)\%$    \\ \hline
2    &  0.729  &  9.20         &    0.176  &   -47.1  \\ 
3    &  0.757  &  13.4         &    0.241  &   -27.8   \\ 
4    &  0.755  &  13.1         &    0.231  &   -30.6  \\ 
5    &  0.754  &  12.9         &    0.267  &   -19.8   \\
6    &  0.756  &  13.2         &    0.257  &   -22.9   \\ 
7    &  0.748  &  12.0         &    0.282  &   -15.5   \\
8    &  0.752  &  12.6         &    0.272  &   -18.4  \\
9    &  0.742  &  11.1         &    0.291  &   -12.7      \\
10   &  0.748  &  11.9         &    0.282  &   -15.5    \\ 
11   &  0.737  &  10.3         &    0.297  &   -10.9  \\ 
12   &  0.743  &  11.3         &    0.289  &   -13.4  \\  
13   &  0.732  &  9.61         &    0.302  &   -9.51  \\ 
14   &  0.739  &  10.7         &    0.294  &   -11.9 \\ \hline
\end{tabular}
\caption{\small
Anharmonic oscillator. Strong-coupling amplitudes and exponents, with the related 
percentage errors, predicted by self-similar factor approximants.}
\end{table}

\begin{table}[hp]
\centering
\renewcommand{\arraystretch}{1.2}
\begin{tabular}{|l|c|c|c|c|} \hline
$k$  &  $B_k$  &  $\ep(B_k)\%$  &  $\nu_k$   &  $\ep(\nu_k)\%$    \\ \hline
2    &  0.727  &  8.87         &    0.300  &   -10.0  \\ 
3    &  0.727  &  8.89         &    0.289  &   -13.2   \\ 
4    &  0.727  &  8.90         &    0.289  &   -13.3  \\ 
5    &  0.713  &  6.78         &    0.310  &   -6.88   \\
6    &  0.712  &  6.57         &    0.312  &   -6.44   \\ 
7    &  0.702  &  5.04         &    0.319  &   -4.32   \\
10   &  0.698  &  4.53         &    0.322  &   -3.42    \\ 
11   &  0.695  &  4.08         &    0.324  &   -2.92  \\ 
13   &  0.690  &  3.23         &    0.326  &   -2.06  \\ 
14   &  0.688  &  3.00         &    0.327  &   -1.85 \\ \hline
\end{tabular}
\caption{\small
Anharmonic oscillator. Strong-coupling amplitudes and exponents, with the related 
percentage errors, predicted by self-similar Borel summation.}
\end{table}

\section{$O(N)$ symmetric $\varphi^4$ field theory}

Now let us consider the problems, where the answer for the strong-coupling limit is 
not known, so that the found results are to be considered as predictions. We start with 
weak-coupling expansions whose coefficients are given with the accuracy not more than 
six digits. Therefore rounding up the numbers, we can set zero the quantities of order 
or less than $10^{-6}$. We shall employ both ways of defining the strong-coupling 
amplitudes and exponents, by directly constructing self-similar factor approximants 
and by using the self-similar Borel summation.

An interesting problem in field theory is the behavior of Gell-Mann-Low functions. 
Let us start with the $O(N)$ symmetric $\varphi^4$ field theory, whose Gell-Mann-Low 
function $\beta(g)$ is a function of the effective coupling parameter 
$g = \lambda/(4\pi)^2$ entering the equation
\be
\label{53}
 \mu\; \frac{\prt g}{\prt\mu} = \bt(g) \;  ,
\ee
where $\mu$ is the renormalization scale. This function can be calculated, within the 
minimal subtraction scheme $\overline{\rm MS}$, by means of perturbation theory yielding
\be
\label{54}
\bt_k(g) = g^2 \sum_{n=0}^k b_n g^n \qquad ( g\ra 0 ) \; .
\ee
The coefficients $b_n$ are known in six-loop approximation \cite{Kompaniets_25} for 
arbitrary $N$ and for $N = 1$, in the seven-loop approximation \cite{Schnetz_26}. In 
Table 6, we give the explicit values of the coefficients for $N = 0,1,2,3,4$. 

\begin{table}[hp]
\centering
\renewcommand{\arraystretch}{1.2}
\begin{tabular}{|c|c|c|c|c|c|} \hline
$N$    &     $0$   &      $1$  &     $2$   &    $3$    &   $4$   \\ \hline
$b_0$  &  2.66667  &  3        &  3.33333  &  3.66667  &  4.0    \\ 
$b_1$  & -4.66667  & -5.66667  & -6.66667  & -7.66667  & -8.66667 \\ 
$b_2$  &  25.4571  &  32.5497  &  39.9478  &  47.6514  &  55.6606 \\ 
$b_3$  & -200.926  & -271.606  & -350.515  & -437.646  & -532.991  \\
$b_4$  &  2003.98  &  2848.57  &  3844.51  &  4998.62  &  6317.66  \\ 
$b_5$  & -23314.7  & -34776.1  & -48999.1  & -66242.7  & -86768.4 \\
$b_6$  &           &  474651   &           &           &  \\ \hline
\end{tabular}
\caption{\small
Coefficients of weak-coupling expansion for the Gell-Mann-Low function of the 
$N$-component $\vp^4$ field theory.}
\end{table}

Employing our method, we find the large-variable amplitudes and exponents for the limit
\be
\label{55}
\bt_k^*(g) \simeq B_k g^{\nu_k} \qquad ( g\ra \infty) \;   ,
\ee
for different $N$ from $0$ to $4$, by the direct self-similar approximation and using 
the self-similar Borel summation. The results are presented in Table 7 ($N = 0$), 
Table 8 ($N=1$), Table 9 ($N=2$), Table 10 ($N=3$), and Table 11 ($N=4$). 

\begin{table}[hp]
\centering
\renewcommand{\arraystretch}{1.2}
\begin{tabular}{|c|c|c|c|c|} \hline
$k$ &  $B_k$  & $\nu_k$  & $B_k(Borel)$ & $\nu_k(Borel)$  \\ \hline
2   &  1.747  &  1.809   &    2.42      &  1.53    \\ 
3   &  1.696  &  1.764   &    1.57      &  1.77   \\ 
4   &  1.699  &  1.769   &    1.42      &  1.83  \\ 
5   &  1.698  &  1.764   &              &    \\ \hline
\end{tabular}
\caption{\small
Strong-coupling amplitudes and exponents for the Gell-Mann-Low function of the 
$N$-component $\vp^4$ field theory, predicted by self-similar factor 
approximants and self-similar Borel summation for $N=0$.}
\end{table}

\begin{table}[hp]
\centering
\renewcommand{\arraystretch}{1.2}
\begin{tabular}{|c|c|c|c|c|} \hline
$k$ &  $B_k$  & $\nu_k$  & $B_k(Borel)$ & $\nu_k(Borel)$  \\ \hline
2   &  1.922  &  1.803   &    2.69      &  1.51    \\ 
3   &  1.856  &  1.750   &    1.84      &  1.72   \\ 
4   &  1.859  &  1.757   &    1.69      &  1.77  \\ 
5   &  1.857  &  1.749   &              &        \\ 
6   &  1.857  &  1.750   &    1.70      &  1.77    \\ \hline
\end{tabular}
\caption{\small
Strong-coupling amplitudes and exponents for the Gell-Mann-Low function of the 
$N$-component $\vp^4$ field theory, predicted by self-similar factor 
approximants and self-similar Borel summation for $N=1$.}
\end{table}

\begin{table}[hp]
\centering
\renewcommand{\arraystretch}{1.2}
\begin{tabular}{|c|c|c|c|c|} \hline
$k$ &  $B_k$  & $\nu_k$  & $B_k(Borel)$ & $\nu_k(Borel)$  \\ \hline
2   &  2.102  &  1.800   &    2.96      &  1.49    \\ 
3   &  2.015  &  1.735   &    2.16      &  1.67   \\ 
4   &  2.020  &  1.744   &    2.03      &  1.71  \\ 
5   &  2.017  &  1.735   &    2.18      &  1.67    \\  \hline
\end{tabular}
\caption{\small
Strong-coupling amplitudes and exponents for the Gell-Mann-Low function of the 
$N$-component $\vp^4$ field theory, predicted by self-similar factor 
approximants and self-similar Borel summation for $N=2$.}
\end{table}

\begin{table}[hp]
\centering
\renewcommand{\arraystretch}{1.2}
\begin{tabular}{|c|c|c|c|c|} \hline
$k$ &  $B_k$  & $\nu_k$  & $B_k(Borel)$ & $\nu_k(Borel)$  \\ \hline
2   &  2.286  &  1.798   &    3.21      &  1.49    \\ 
3   &  2.175  &  1.718   &    2.49      &  1.63   \\ 
4   &  2.182  &  1.732   &    2.93      &  1.66  \\ 
5   &  2.178  &  1.719   &    2.50      &  1.63    \\  \hline
\end{tabular}
\caption{\small
Strong-coupling amplitudes and exponents for the Gell-Mann-Low function of the 
$N$-component $\vp^4$ field theory, predicted by self-similar factor 
approximants and self-similar Borel summation for $N=3$.}
\end{table}

\begin{table}[hp]
\centering
\renewcommand{\arraystretch}{1.2}
\begin{tabular}{|c|c|c|c|c|} \hline
$k$ &  $B_k$  & $\nu_k$  & $B_k(Borel)$ & $\nu_k(Borel)$  \\ \hline
2   &  2.474  &  1.797   &    3.45      &  1.49    \\ 
3   &  2.336  &  1.701   &    2.82      &  1.60   \\ 
4   &  2.345  &  1.721   &    2.74      &  1.62  \\ 
5   &  2.340  &  1.702   &    2.81      &  1.61    \\  \hline
\end{tabular}
\caption{\small
Strong-coupling amplitudes and exponents for the Gell-Mann-Low function of the 
$N$-component $\vp^4$ field theory, predicted by self-similar factor 
approximants and self-similar Borel summation for $N=4$.}
\end{table}

In Tables $7$ to $12$, we give the predicted values for the strong-coupling 
amplitudes and exponents of the Gell-Mann-Low functions. Since the exact 
behavior of the Gell-Mann-Low functions in the strong-coupling limit is not 
known, the actual errors of the predicted values cannot be defined. What one 
could do is to compare our results with calculations accomplished by other 
methods. 

Estimates for the strong-coupling exponent $\nu$ of the Gell-Mann-Low function of the
$N$-component $\vp^4$ field theory have been done by other methods only for $N = 1$. 
Thus, Borel-type summation with conformal mapping gives \cite{Kazakov_27,Chetyrkin_28} 
the estimates around $\nu =2$, although this value essentially depends on the type of 
the Borel transform and on the used conformal mapping. Nevertheless, it is in agreement 
with our prediction. 
 
The large-variable exponent can also be estimated by resorting to optimized perturbation 
theory based on the introduction of control functions or parameters in perturbative series 
and defining these parameters by some optimization conditions, like minimal difference or 
minimal derivative conditions. This approach, advanced in \cite{Yukalov_29,Yukalov_30}, 
has been used for numerous problems (see reviews \cite{Yukalov_15,Yukalov_16}), including 
quantum field theory 
(e.g. \cite{Krein_31,Kneur_32,Kneur_33,Kneur_34,Duarte_35,Rosa_36,Duarte_37}). The 
strong-coupling behavior of the Gell-Mann-Low function for $\varphi^4$ theory with $N=1$ 
is estimated in \cite{Sissakian_38}, were it is found that for $g \ra \infty$, in the 
second order of perturbation theory, one gets $\nu \approx 1.5$, which agrees with the 
second order approximation in Table 8.

\section{Quantum electrodynamics}

The Gell-Mann-Low function $\beta(\alpha)$ in quantum electrodynamics is a function of 
the coupling parameter $\alpha$ in the $\overline{\rm MS}$ scheme, satisfying the equation
\be
\label{56}
\mu^2\; \frac{\prt }{\prt\mu^2}\;\left(\frac{\al}{\pi}\right)  = \bt(\al) \; . 
\ee
The weak-coupling expansion, in five-loop approximation, taking into account the electron, 
but neglecting the contributions of leptons with higher masses, that is, muons and tau 
leptons, reads as
\be
\label{57}
\bt_k(\al) = \left(\frac{\al}{\pi}\right)^2 \sum_{n=0}^k b_n 
\left(\frac{\al}{\pi}\right)^n \qquad ( \al \ra 0 ) \; ,
\ee
with the coefficients \cite{Kataev_39}
$$
b_0 = \frac{1}{3} \; , \qquad b_1 = \frac{1}{4} \; , \qquad
b_2 = -0.107639 \; , 
$$
$$
b_3 = - 0.523614 \; , \qquad b_4 = 1.47072 \;  .
$$
For the large-coupling behavior 
\be
\label{58}
 \bt_k^*(\al) \simeq B_k \left(\frac{\al}{\pi}\right)^{\nu_k} \qquad
(\al \ra \infty ) \;  ,
\ee
we find the amplitude and exponent shown in Table 12. Notice that the large-variable 
exponents diminish with increasing order. 

\begin{table}[hp]
\centering
\renewcommand{\arraystretch}{1.2}
\begin{tabular}{|c|c|c|c|c|} \hline
$k$ &  $B_k$  & $\nu_k$  & $B_k(Borel)$ & $\nu_k(Borel)$  \\ \hline
2   &  0.416  &  2.466   &    0.333      &  2.64    \\ 
3   &  0.469  &  2.167   &    0.587      &  2.12   \\ 
4   &  0.476  &  2.096   &    1.560      &  1.64    \\  \hline
\end{tabular}
\caption{\small
Strong-coupling amplitudes and exponents for the Gell-Mann-Low function of quantum 
electrodynamics, predicted by self-similar factor approximants and self-similar 
Borel summation.}
\end{table}

It is useful to note that the Borel summation is not necessarily preferable than 
the direct self-similar approximation. The Borel summation is supposed to work well, 
when the coefficients of the asymptotic series factorially grow, while the Borel
transform compensates this growth. However, when the coefficients $b_n$ of the 
perturbative series do not exhibit persistent growth close to factorial, but vary 
rather chaotically, Borel summation can result in sharp variations of the calculated
quantity. This happens for the series representing the Gel-Mann-Low function of
quantum electrodynamics. In the case of factorial growth, the ratio $|b_{n+1}/(n+1)b_n|$
would be approximately constant, but here we have rather different abruptly varying
ratios $|b_2/2 b_1| = 0.215$, but $|b_3/3 b_2| = 1.622$, and $|b_4/4 b_3| = 0.702$. 
Because of this, the Borel summation in different orders essentially varies, as is 
seen from Table 12. On the contrary, the direct self-similar approximation is not so 
sensible to the coefficient variation, which results in more smooth changes of the 
quantities in different orders. 

Self-similar factor approximants do not strongly depend on the used scheme. Thus for the 
coefficients $b_n$, calculated \cite{Kataev_39} in the on-shell scheme, employing the 
self-similar factor approximants, we get $B = 0.43$ and $\nu = 2.19$, while in the momentum 
subtraction scheme, $B = 0.49$ and $\nu = 2.31$. 

Since the analytic expression for the whole $\beta$ function can be explicitly 
represented by self-similar factor approximants (\ref{18}), it is possible to solve 
the equation for the running coupling (\ref{56}), taking for the boundary condition 
the value $\alpha(m_Z) = 0.007815$ at the Z-boson mass $m_Z = 91.1876$ GeV. The 
resulting running coupling grows from zero to divergence at $\mu_0 \approx 10^{261}$ GeV 
by the law 
\be
\label{59}
 \al \simeq \frac{2.74}{[\; \ln(\mu_0/\mu) \;]^{0.68} } \qquad
(\mu \ra \mu_0 - 0 ) \;  .
\ee
This value of $\mu_0$ is incomparably larger than the Landau pole that is of the order 
of $10^{30}-10^{40}$ GeV \cite{Gockeler_40,Deur_40}. Because of the so large value of 
the divergence point $\mu_0 \approx 10^{261}$ GeV, it is hardly achievable. Also notice 
that, with the increasing order of the approximants, the pole shifts closer to infinity.   

The self-similar factor approximants, as has been checked by numerous examples, do 
possess the power of well extrapolating perturbative series up to large-variable limit,
including the variable tending to infinity, provided the considered perturbative expansion
is correctly derived taking into account all physics of the problem, even if the influence
of some effects seems to be negligible for asymptotically small variables. Therefore the 
occurrence of the divergence of the running coupling at a finite value of the scale 
$\mu_0$ is not caused by the extrapolation as such. To our mind, this is the result of
incomplete account of physical effects at small couplings, which leads to the loss of
information and to not completely correct coefficients of the perturbative expansions. 
Self-similar approximation theory, by its construction, has the ability of extracting 
the information contained in the perturbative coefficients, even when the influence of 
some effects is small, and allows for the self-similar extrapolation of this information
to large variables, such as coupling. However if some physical effects have been neglected 
in the derivation of a perturbative expansion, then the expansion coefficients contain
no information on them, hence there is nothing to be extrapolated. For instance, in the 
derivation of the weak-coupling expansion (\ref{57}), the contributions of leptons with 
higher masses, that is, muons and tau leptons, have been neglected, hence the coefficients
$b_n$ contain no information on the effects that can become important at large coupling. 
Not correctly defined coefficients can lead to the appearance of unphysical divergences.

\section{Quantum chromodynamics}

The running quark-gluon coupling $\alpha_s$, as a function of the renormalization scale 
$\mu$, satisfies the equation
\be
\label{60}
\mu^2\; \frac{\prt\al_s }{\prt\mu^2}  = \bt(\al_s) \;   .
\ee
Keeping in mind the realistic case of three colors ($N_c = 3$), the beta function
can be written \cite{Deur_41} as the weak-coupling expansion
\be
\label{61}
\bt_k(\al_s) = -\; \frac{\al_s^2}{4\pi} \sum_{n=0}^k b_n 
\left(\frac{\al_s}{4\pi}\right)^n \qquad ( \al_s \ra 0 ) \;   .
\ee
In the five-loop approximation, with the $\overline{\rm MS}$ renormalization scheme, the 
coefficients are \cite{Luthe_42,Baikov_43,Herzog_44}
$$
b_0 = 11 -\; \frac{2}{3}\; n_f \; , \qquad b_1 = 102 -\;\frac{38}{3} \; n_f \; ,
\qquad
b_2 = \frac{2857}{2} \; - \; \frac{5033}{18}\; n_f \; + \frac{325}{54}\; n_f^2 \; ,
$$
$$
b_3 = \frac{149 753}{6} + 3564\zeta_3 - \left( \frac{1078361}{162} +
\frac{6508}{27}\zeta_3\right) \; n_f \; + \;
\left( \frac{50065}{162} + \frac{6472}{81}\zeta_3\right) \; n_f^2 \;  + \; 
\frac{1093}{729}\; n_f^3 \; ,
$$
$$
b_4 = \frac{8157455}{16} + \frac{621885}{2}\; \zeta_3 -\;
\frac{88209}{2}\; \zeta_4 - 288090 \; \zeta_5 \; +
$$
$$
+ \;
\left( -\; \frac{336460813}{1944} -\;  \frac{4811164}{81}\zeta_3
+ \frac{33935}{6}\; \zeta_4 + \frac{1358995}{27}\;\zeta_5
\right) \; n_f \; +
$$
$$
+ \; 
\left( \frac{25960913}{1944} + \frac{698531}{81}\zeta_3 - \;
\frac{10526}{9}\; \zeta_4 -\;  \frac{381760}{81} \;\zeta_5
\right) \; n_f^2 \; + 
$$
\be
\label{62}
+  \; 
\left( -\; \frac{630559}{5832} -\; \frac{48722}{243}\zeta_3 +
\frac{1618}{27}\; \zeta_4 + \frac{460}{9} \;\zeta_5 \right) \; n_f^3 \; + \; 
\left( \frac{1205}{2916} -\; \frac{152}{81}\; \zeta_3\right) n_f^4 \; ,
\ee
where $\zeta_n$ is the Riemann zeta function. For the physically realistic case of six 
flavors ($n_f = 6$), we have
$$
b_0 = 7 \; , \qquad b_1 = 26 \; , \qquad b_2 = -32.5 \; , \qquad
b_3 = 2472.28 \qquad b_4 = -11783.7 \;   .
$$
      
Using the self-similar factor approximants gives in second order
\be
\label{63}
 \bt_2^*(\al_s) = - \; \frac{7}{4\pi} \; \al_s^2 ( 1 + A_1 \al_s)^{n_1} \; ,
\ee
where the control parameters $A_1$ and $n_1$ are found from the training conditions 
(\ref{22}), yielding
$$
 A_1 = 0.494517 \; , \qquad n_1 = 0.597701 \;  .
$$
In third order, the beta function reads as
\be
\label{64}
 \bt_3^*(\al_s) = - \; \frac{7}{4\pi} \; \al_s^2 ( 1 + A_1 \al_s)^{n_1}
(1 + A_2 \al_s)^{n_2} \;   .
\ee
The control parameters $A_1$, $n_1$, and $n_2$ are prescribed by the training conditions
(22), as is described in Sec. 3, and $A_2$ is treated as an additional control parameter 
defined by the minimal-derivative optimization condition $|\prt\nu_3/\prt A_2|=\dlt$, 
where $\dlt= 10^{-6}$, as is explained in Appendix. Thus we obtain
$$
 A_1 = 0.490298 \; , \qquad n_1 = 0.602838 \; , \qquad
 A_2 = 412.291 \; , \qquad n_2 = 0.734531 \times 10^{-8} \;  .
$$

This leads to the strong-coupling limit
\be
\label{65}
\bt_2^*(\al_s) \simeq - 0.366 \; \al_s^{2.598} \; , \qquad
\bt_3^*(\al_s) \simeq - 0.362 \; \al_s^{2.603}  \qquad 
( \al_s \ra \infty) \;  .
\ee
   
Similarly, employing the self-similar Borel summation of Sec. 5, we get the strong-coupling 
behavior
\be
\label{66}
\bt_2^*(\al_s) \simeq - 0.255 \; \al_s^{2.748} \; , \qquad
\bt_3^*(\al_s) \simeq - 0.254 \; \al_s^{2.751}  \qquad 
( \al_s \ra \infty) \;    .
\ee

As is mentioned in Appendix, self-similar approximants are uniquely defined for the
functions in the complex range. However, when we are interested only in real-valued 
functions, we have to discard occasionally appearing complex-valued approximants. 
In the case of the Gell-Mann-Low function of quantum chromodynamics, the fourth order 
approximant is complex and is discarded. So, we deal with the second and third orders 
that are close to each other.

Having in hands the beta function, it is straightforward to solve equation (\ref{60})
for the running coupling. As a boundary condition, we can take the value  
$\alpha(m_Z) = 0.1181$ at the Z-boson mass $m_Z = 91.1876$ GeV. The running coupling,
with decreasing $\mu$, grows from zero to the logarithmic divergence 
\be
\label{67}
\al_s \simeq \frac{0.91}{[\;\ln(\mu/\mu_c)\;]^{0.63} } \qquad
(\mu \ra \mu_c + 0 )
\ee
at the point $\mu_c = 0.1$ GeV, which is an order lower than the Landau pole being 
about $1$ GeV \cite{Deur_41}. 

Similarly to quantum electrodynamics, the appearance of the divergence of the running 
coupling is not caused by the extrapolation as such, but it can be induced by the 
absence of information on physical effects that could become important at large couplings,
while they where completely ignored in the derivation of the expansion (\ref{61}).
For example, the possibility of arising bound states is ignored in the derivation of 
expansion (\ref{61}), which makes the coefficients $b_n$ incompletely defined, thus
making impossible the precise extrapolation to large couplings. Self-similar approximation
theory is able to extrapolate functions to arbitrary values of their variables, provided
the perturbative expansion contains information for being extrapolated. When there is 
no information, there is nothing to be extrapolated, and unphysical divergences can happen. 
It is not the method of extrapolation that is guilty but incorrect expansion coefficients.

\section{Conclusion}

We have addressed an important but rather complicated problem of the possibility to 
reconstruct the whole functions being based only on their asymptotic expansions over 
coupling parameters or other variables. Of special interest is the problem of finding 
the strong-coupling or large-variable behavior of physical quantities from the knowledge 
of only several terms of their weak-coupling or small-variable expansions. The method 
is presented allowing for resolving this problem. The approach is based on the ideas 
of renormalization group theory, approximation theory, dynamical theory, and optimal 
control theory. 

The method allows for the extrapolation of small-variable expansions because the 
coefficients of such expansions contain information on the whole functions to be 
restored. This information can be extracted by observing self-similarity between the
subsequent approximations and considering the passage from one approximation to another
as the motion in the approximation space, with the order of approximation playing the 
role of discrete time. The sequence of approximations can be transformed to a 
dynamical system called approximation cascade. The fixed point of the cascade represents
the effective limit of the approximation sequence. The extrapolation is efficient even 
for the most difficult case of finding large-variable or strong-coupling limits. 
      
Convergence and accuracy of the method are illustrated by explicit examples, including 
the so-called zero-dimensional field theory and anharmonic oscillator, which serve as
typical touchstones for testing summation methods of divergent series. Strong-coupling 
behavior of Gell-Mann-Low functions in multicomponent field theory, quantum 
electrodynamics, and quantum chromodynamics is found, being based on their weak-coupling 
perturbative expansions.

\appendix
\section{Appendix. Properties of self-similar factor approximants}

\numberwithin{equation}{section}
\setcounter{equation}{0}

Self-similar factor approximants
\be
\label{A1}
f_k^*(x) = \prod_{j=1}^{N_k} ( 1 + A_j x)^{n_j} \;   ,
\ee
generally, can be considered on the complex plane \cite{Yukalov} of the variable 
$x \in \mathbb{C}$, although in applications one is more often interested in real 
variables $x \in \mathbb{R}_+ =  [0,\infty)$. The form (\ref{A1}) is real-valued 
provided that $x \in \mathbb{R}_+$ and either $A_j x > -1$, with $A_j$ and $n_j$ 
being real, or in the product (\ref{A1}) there occur complex conjugate pairs of $A_j$ 
and $n_j$, so that the product of two factors is real, where
$$
\left| \; ( 1 + A_j x)^{n_j} \; \right|^2 = 
| \; 1 + A_j x \; |^{2{\rm Re}\;n_j} 
\exp\{ - 2 {\rm Im} (n_j) {\rm arg}(1 + A_j x) \} \;   .
$$
The control parameters $A_j$ and $n_j$ are defined by the training equations (\ref{21}).
When we are looking for a real-valued function, occasional complex-valued solutions are 
discarded. 

Considering the sequence of the terms $\{f_k^*(x)\}$, since $N_k \ra \infty$ as 
$k \ra \infty$, one comes to the limiting form
\be
\label{A2}
f^*(x) = \prod_{j=1}^\infty ( 1 + A_j x)^{n_j} \;   .
\ee

\vskip 2mm

{\bf Definition 1}. The sequence $\{f_k^*(x)\}$ converges to $f^*(x)$ at a point $x$ if
there are not more than a finite number of zero factors $(1 + A_j x)$ and
\be
\label{A3}
 0 \leq |\; f^*(x) \; | < \infty \;  .
\ee 
If the sequence $\{f_k^*(x)\}$ converges at all points $x$ of a given domain 
$\mathbb{D} \subset \mathbb{R}_+$, then one says that it uniformly converges on the 
given domain. 

\vskip 2mm
{\bf Cauchy criterion}. The sequence $\{f_k^*(x)\}$ converges if and only if for any 
$\varepsilon > 0$ there is $j_\varepsilon > 0$ such that 
\be
\label{A4}
\left| \; \prod_{j=j_\ep}^m ( 1 + A_j x)^{n_j} - 1 \; \right| < \ep   
\ee
for all $m > j_\varepsilon$.

\vskip 2mm
{\bf Theorem 1}. The sequence $\{f_k^*(x)\}$ converges if and only if the sequence of
the sums
\be
\label{A5}
S_k(x) = \sum_{j=1}^{N_k} n_j \ln ( 1 + A_j x )
\ee
converges. 

\vskip 2mm
{\it Proof}. The proof is clear from the identity
$$  
f_k^*(x) = \exp\{ \ln f_k^*(x) \} = e^{S_k(x)} \;   .
$$

\vskip 2mm
{\bf Corollary 1}. From the Cauchy criterion and Theorem 1 it follows that
\be
\label{A6}
  \lim_{j\ra\infty} ( 1 + A_j x)^{n_j} = 1 
\ee
and
\be
\label{A7}
 \lim_{j\ra\infty} n_j \ln ( 1 + A_j x) = 0\;  .
\ee
This implies that either $A_j \ra 0$ or $n_j \ra 0$. In any case
\be
\label{A8}
\lim_{j\ra\infty} n_j A_j = 0\; .
\ee

\vskip 2mm
{\bf Theorem 2}. If the sequence of the terms 
\be
\label{A9}
f_k(x) = \sum_{n=0}^k a_n x^n
\ee
converges on a domain $\mathbb{D} \subset  \mathbb{R}_+$ to a function
\be
\label{A10}
 f(x) = \sum_{n=0}^\infty a_n x^n \;  ,
\ee
then the sequence $\{f_k^*(x)\}$ converges on the same domain to $f^*(x)$ coinciding 
with $f(x)$,
\be
\label{A11}
f^*(x) = f(x) \; .
\ee

\vskip 2mm
{\it Proof}. The proof is given in Ref. \cite{Yukalov_45}.

\vskip 2mm
{\bf Theorem 3}. If $A_j x \geq -1$ and the sequence of the sums
\be
\label{A12}
s_k = \sum_{j=1}^{N_k} |\; n_j A_j \; |
\ee
converges, then the sequence of the approximants $\{f_k^*(x)\}$ converges. 

\vskip 2mm
{\it Proof}. For the approximants $f_k^*(x)$, we have
$$
|\; f_k^*(x) \; | \leq \exp\{ |\; S_k(x) \; | \} \;  ,
$$
where
$$
|\; S_k(x) \; | \leq \sum_{j=1}^{N_k} |\; n_j \ln ( 1 + A_j x) \; | \;  .
$$
Since 
$$
\ln ( 1 + z ) \leq z \qquad ( z \geq - 1) \;  ,
$$
then
$$
|\; S_k(x) \; | \leq \sum_{j=1}^{N_k} | \; n_j A_j \; | x = s_k x \;  .
$$
Therefore
$$
|\; f_k^*(x) \; | \leq e^{s_k x} \;   .
$$
The sequence $\{s_k\}$, by assumption converges, hence $\{\exp(s_k x)\}$ converges, 
because of which $f_k^*(x)$ converges. 

\vskip 2mm
{\bf Corollary 2}. If the sequence $\{s_k\}$ converges, then the scaling 
\be
\label{A13}
\overline A_j = \lbd_j A_j \; , \qquad \overline n_j = \frac{n_j}{\lbd_j} 
\ee
does not change the convergence of $s_k$, as far as $n_j A_j=\bar{n}_j \bar{A}_j$. 

\vskip 2mm
Thus, when one of the parameters $A_j$ is not defined, as it happens for odd approximants,
it is possible to fix one of $A_j$ by treating it as a control parameter prescribed by one 
of known optimization conditions \cite{Yukalov_15,Yukalov_16}. For instance, being the most 
interested in the exponent of the large-variable behavior
\be
\label{A14}
 f_k^*(x) \simeq B_k x^{\nu_k} \qquad ( x\ra \infty) \;  ,
\ee
in which 
\be 
\label{A15}
 B_k = \prod_{j=1}^{N_k} A_j^{n_j} \; , \qquad 
\nu_k = \sum_{j=1}^{N_k} n_j \;  ,
\ee
it is possible to resort to the minimal-derivative condition with respect to 
$\partial \nu_k / \partial A_j$, with a fixed $j$. Very often, the solution to the 
equation $\partial \nu_k / \partial A_j = 0$ does not exist, then it is sufficient to
employ the optimization condition 
\be
\label{A16}
 \min \left| \; \frac{\prt\nu_k}{\prt A_j} \; \right| = \dlt \;  ,
\ee
where $\delta$ is the required accuracy, say $\delta = 10^{-6}$. However, condition
(\ref{A16}) makes calculations cumbersome, this is why it can be used only in low orders
and when the method of determining the exponent $\nu_k$ by means of the diff-log 
transformation of Sec. 4 does not have real-valued solutions. 
  
For realistic complicated problems, no general form of the coefficients in the 
weak-coupling expansions are usually known, but merely a few terms of the expansion
are available. Therefore the general convergence properties cannot be checked, but
only numerical convergence can be observed, when the subsequent approximations are
close to each other.

\section*{Declaration of competing interest}

The authors declare that they have no known competing financial interests or personal 
relationships that could have appeared to influence the work reported in this paper.

\section*{Author contributions}

{\bf V.I. Yukalov}: Conceptualization, Methodology, Writing - Original Draft, 
Writing - Review $\&$ Editing
{\bf E.P. Yukalova}: Software, Formal analysis, Writing - Review $\&$ Editing

\end{document}